\documentclass[useAMS,usenatbib,usegraphicx]{mn2e}

\title[Assessing potential cluster Cepheids]{Assessing potential cluster Cepheids from a new distance and reddening parameterization and 2MASS photometry}
\author[Majaess, Turner and Lane]
{D.~J. Majaess$^{1,3,4}$\thanks{Email: dmajaess@ap.smu.ca}, D.~G. Turner$^{1,3,4}$ and D.~J. Lane$^{1,2}$ \\ \\
  $^1$Department of Astronomy and Physics, Saint Mary's University,
    Halifax, Nova Scotia B3H 3C3, Canada \\
  $^2$The Abbey Ridge Observatory, Stillwater Lake, Nova Scotia, Canada \\
  $^3$Visiting Astronomer, Dominion Astrophysical Observatory, Herzberg Institute of Astrophysics,  
    National Research Council of Canada \\
  $^4$Visiting Astronomer, Harvard College Observatory Photographic Plate Stacks 
}
\date{Accepted 2008 August 13. Received 2008 July 11; in original form 2008 April 27}
\pagerange{\pageref{firstpage}--\pageref{lastpage}} \pubyear{2008}

\begin{document}

\label{firstpage}

\maketitle

\begin{abstract}
A framework is outlined to assess Cepheids as potential cluster members from readily available photometric observations. A relationship is derived to estimate colour excess and distance for individual Cepheids through a calibration involving recently published HST parallaxes and a cleaned sample of established cluster Cepheids. Photometric {\it V--J} colour is found to be a viable parameter for approximating a Cepheid's reddening. The non-universal nature of the slope of the Cepheid PL relation for {\it BV} photometry is confirmed. By comparison, the slopes of the {\it VJ} and {\it VI} relations seem relatively unaffected by metallicity. A new Galactic Cepheid confirmed here, GSC 03729-01127 (F6-G1 Ib), is sufficiently coincident with the coronal regions of Tombaugh 5 to warrant follow-up radial velocity measures to assess membership. CCD photometry and O--C diagrams are presented for GSC 03729-01127 and the suspected cluster Cepheids AB Cam and BD Cas. Fourier analysis of the photometry for BD Cas and recent estimates of its metallicity constrain it to be a Population I overtone pulsator rather than a Type II s-Cepheid. AB Cam and BD Cas are not physically associated with the spatially-adjacent open clusters Tombaugh 5 and King 13, respectively, the latter being much older ($\log \tau \simeq 9$) than believed previously. Rates of period change are determined for the three Cepheids from archival and published data. GSC 03729-01127 and AB Cam exhibit period increases, implying fifth and third crossings of the instability strip, respectively, while BD Cas exhibits a period decrease, indicating a second crossing, with possible superposed trends unrelated to binarity.  More importantly, the observed rates of period change confirm theoretical predictions. The challenges and prospects for future work in this area of research are discussed.
\end{abstract}


\section{Introduction}
The {\it All Sky Automated Survey} \citep[ASAS,][]{po00}, the {\it Northern Sky Variability Survey} \citep[NSVS,][]{wo04}, and {\it The Amateur Sky Survey} \citep[TASS,][]{dr06} have detected many new Cepheid variables through their photometric signatures, resulting in a valuable expansion of the Galactic Cepheid sample \citep{sd04} once confirmed by spectroscopic observation. In the case of GSC 03729-01127 (TASSIV 6349369), a suspected Cepheid studied here, the variable may be an open cluster member and a potentially valuable calibrator for the Cepheid period-luminosity (PL) relation. Cepheids continue to provide the foundation for the universal distance scale, and such variables could serve as an efficient means of quantifying the extinction to Galactic and extragalactic targets.

The results of the seminal Hubble Space Telescope (HST) Key Project yielded a Hubble constant of $H_0=72\pm8$ km s$^{-1}$ Mpc$^{-1}$ \citep{fr01}, a value supported by cosmological constraints inferred from WMAP observations \citep{sp07}. The HST results are tied to Large Magellanic Cloud (LMC) Cepheids, which advantageously provide a large sample of common distance. However, distance estimates for the LMC exhibit an unsatisfactorly large scatter \citep{fr01,be02}, resulting in an uncertain zero-point. Moreover there exists a difference in metallicity between LMC Cepheid variables relative to both Galactic Cepheids and those in galaxies used for calibrating secondary distance candles, the effects of which remain actively debated. \citet{ta03}, for example, suggest that the LMC Cepheid PL relation appears to characterize short-period Cepheids as too bright relative to their Galactic counterparts, and long-period Cepheids as too faint. Conversely, \citet{vl07} and \citet{be07} suggest, on the basis of revised Hipparcos and newly-derived HST parallaxes, that the slopes of the PL relations ($V,I$) for Cepheids in the Galaxy and the LMC are consistent to within their cited uncertainties. Indeed, the results presented in section \ref{extragal}, complementing in part those of \citet{fo07}, appear to confirm ideas put forth in each of the above studies, namely that the slope of the PL relation is not universal in certain passbands, and for {\it VJ} and {\it VI} constructed relations, any putative difference in slope arising from metallicity effects appears negligible in comparison with other concerns and uncertainties related to extragalactic observations. Nevertheless, a consensus has yet to emerge and a resolution to the above debate may be assisted by renewed efforts towards establishing Galactic Cepheids as cluster members, a connection that provides direct constraints on Cepheid luminosities, intrinsic colours, masses, metallicities, and pulsation modes.  

\citet{tu02} compiled an extensive list of suspected cluster Cepheids based upon preliminary analyses, but only a few cluster/Cepheid pairs have been studied with the necessary detail to determine the parameters of the associated clusters accurately, or to obtain the necessary radial velocity measures needed to establish membership in cases where reliable proper motions are unavailable. Efforts to discover new Galactic open clusters \citep{al03,mo03,kro06} and Cepheid variables \citep{po00,wo04,dr06} have resulted in a welcome increase to the number of suspected cluster Cepheids. This paper outlines a framework to assess the viability of such cases efficiently, with an intent to highlight cases requiring further attention and focus. Section \ref{framework} develops a relationship to estimate colour excesses and distances for individual Cepheids from several photometric parameters. Section \ref{cepheids} presents CCD photometry, spectroscopic results, and O--C analyses for the suspected cluster Cepheids BD Cas \citep{ts66,tu02}, AB Cam \citep{vb57,ts66}, and a new Galactic Cepheid confirmed here, GSC 03729-01127. Distances, colour excesses, and ages are also derived for the associated open clusters Tombaugh 5 and King 13 from 2MASS photometry \citep{cu03}.  

\section{Data Acquisition and Reductions}

Light curves for the Cepheids presented here were constructed from CCD photometry obtained with the 0.3-m Schmidt-Cassegrain telescope of the Abbey-Ridge Observatory (ARO), an automated facility located outside of Halifax, Nova Scotia. A description of the facility, its equipment and the data reduction procedures used for the observations are given elsewhere \citep{ma08}. Low dispersion spectra at 120 {\AA} mm$^{-1}$ were obtained for the Cepheids with the Dominion Astrophysical Observatory's 1.8-m Plaskett telescope in October 2006 using the SITe-2 CCD detector. The spectra were reduced and analyzed using the NOAO's routines in \textsc{iraf}, along with software packages by Christian Buil (\textsc{iris}), Valerie Desnoux (\textsc{vspec}) and Robert H. Nelson (\textsc{raverec}). 

Rates of period change for the Cepheids studied here were determined through O--C analyses using light curves constructed primarily from data derived from visual scanning of images in the Harvard College Observatory Photographic Plate Collection. Software tailored for the analysis of Cepheid light curves \citep[see][]{tu04} was used to determine offsets in phase and magnitude space that minimize the $\chi^2$ statistic when matching input light curves to a standard template. 

\setcounter{table}{0}
\begin{table}
\caption[]{Calibrating Data Set.}
\label{calset}
\begin{tabular}{@{\extracolsep{-3mm}}lclccl}
\hline
Cepheid &Period &Cluster &Distance & $E_{B-V}$ &Source \\
&(days) & &(pc) \\
\hline
EV Sct &4.39 &NGC 6664 &1612 &0.64 &(1) \\
CF Cas &4.88 &NGC 7790 &2884 &0.57 &(2,3) \\
CV Mon &5.38 &van den Bergh 1 &1650 &0.75 &(4) \\
QZ Nor &5.47 &NGC 6067 &1621 &0.35 &(5) \\
V Cen &5.49 &NGC 5662 &790 &0.31 &(6) \\
V367 Sct &6.29 &NGC 6649 &1650 &1.27 &(7,8) \\
U Sgr &6.75 &IC 4725 &599 &0.43	&(9,10) \\
DL Cas &8.00 &NGC 129 &1670 &0.47 &(11) \\
S Nor &9.75 &NGC 6087 &902 &0.17 &(12) \\
TW Nor &10.79 &Lyng\aa\ 6 &1923 &1.22 &(13,14,10) \\
V340 Nor &11.29 &NGC 6067 &1621 &0.35 &(15) \\
\\
RT Aur &3.73 &... &417 &0.051 &(16) \\
T Vul &4.44 &... &526 &0.064 &(16) \\
FF Aql &4.47 &... &356 &0.224 &(16) \\
$\delta$ Cep &5.37 &... &273 &0.092 &(16) \\
Y Sge &5.77 &... &469 &0.205 &(16) \\
X Sgr &7.01 &... &333 &0.197 &(16) \\
W Sgr &7.59 &... &439 &0.111 &(16) \\
$\beta$ Dor &9.84 &... &318 &0.044 &(16) \\
$\zeta$ Gem &10.15 &... &360 &0.018 &(16) \\
$\ell$ Car &35.55 &... &498 &0.17 &(16) \\
\hline
\end{tabular}
Data sources: (1) \citet{tu76}, (2) \citet{pe84}, (3) \citet{ta88}, (4) \citet{te98}, (5) \citet{wa85b}, (6) \citet{cl91}, (7) \citet{mv75}, (8) \citet{tu81}, (9) \citet{pe85}, (10) \citet{tu02}, (11) \citet{tu92}, (12) \citet{tu86}, (13) \citet{ma75}, (14) \citet{wa85a}, (15) \citet{wa85b}, (16) \citet{be07}. 
\end{table}

\section{Cepheid-Distance Relation}
\label{framework}
The distance to a Cepheid can be established through adoption of intrinsic parameters from published PL ($\langle M_V\rangle$ versus $\log P$) and period-colour [$(B-V)_0$ versus $\log P$] relations, although such estimates typically idealize the result to an object located near the centre of the instability strip. Neglect of the intrinsic scatter inherent to such relationships can affect estimates of colour excess and distance to individual Cepheids made from them, since the variable may lie anywhere within the strip (e.g., towards the red or blue edge). Large calibrating data sets are therefore required when constructing strict two-parameter Cepheid relations to ensure a reasonably even sampling of both sides of the strip and to avoid a least-squares solution biased towards predominantly red or blue edge objects, an important consideration that is often overlooked. 

\begin{figure}
\begin{center}
\includegraphics[width=7cm]{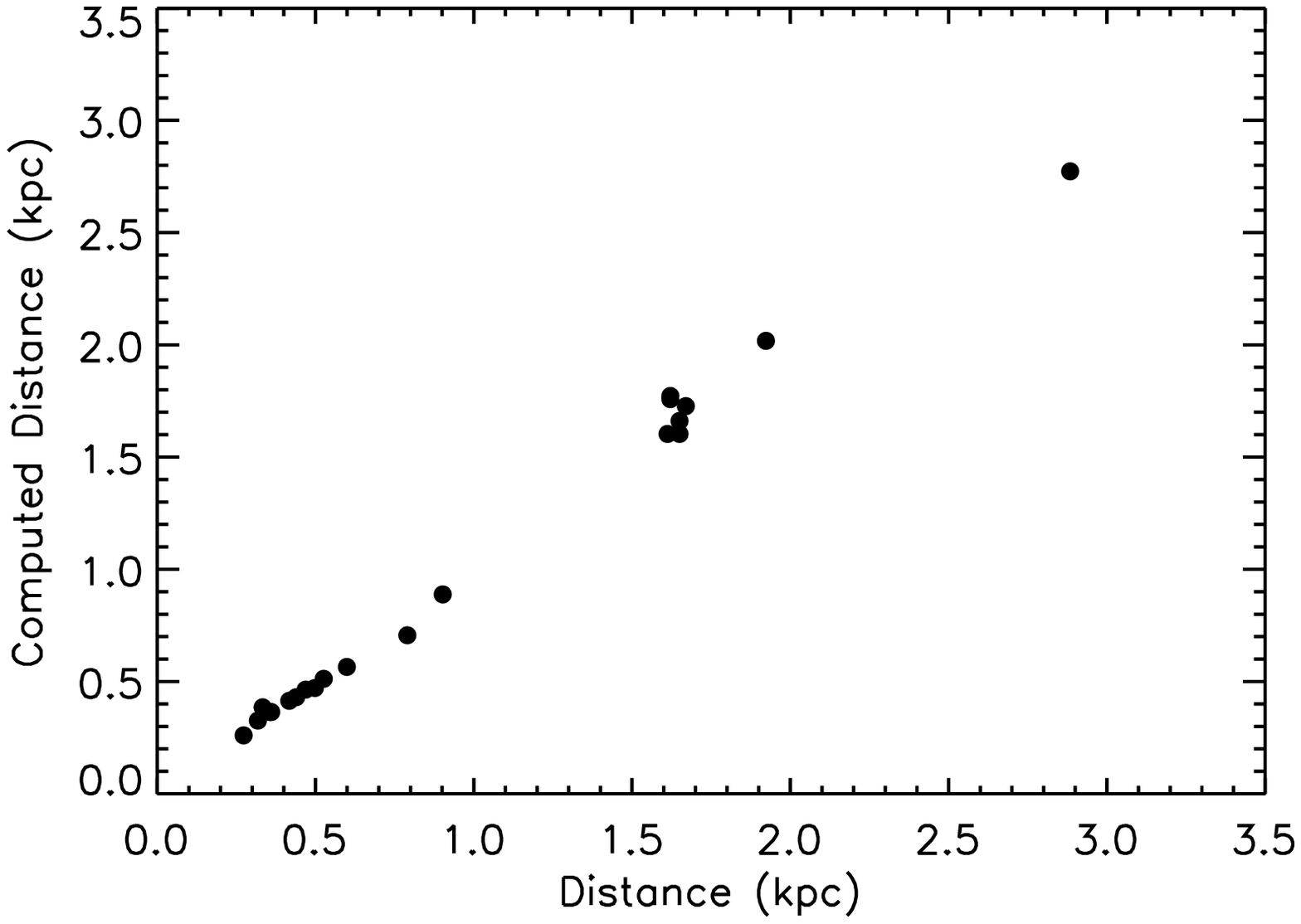}
\includegraphics[width=7cm]{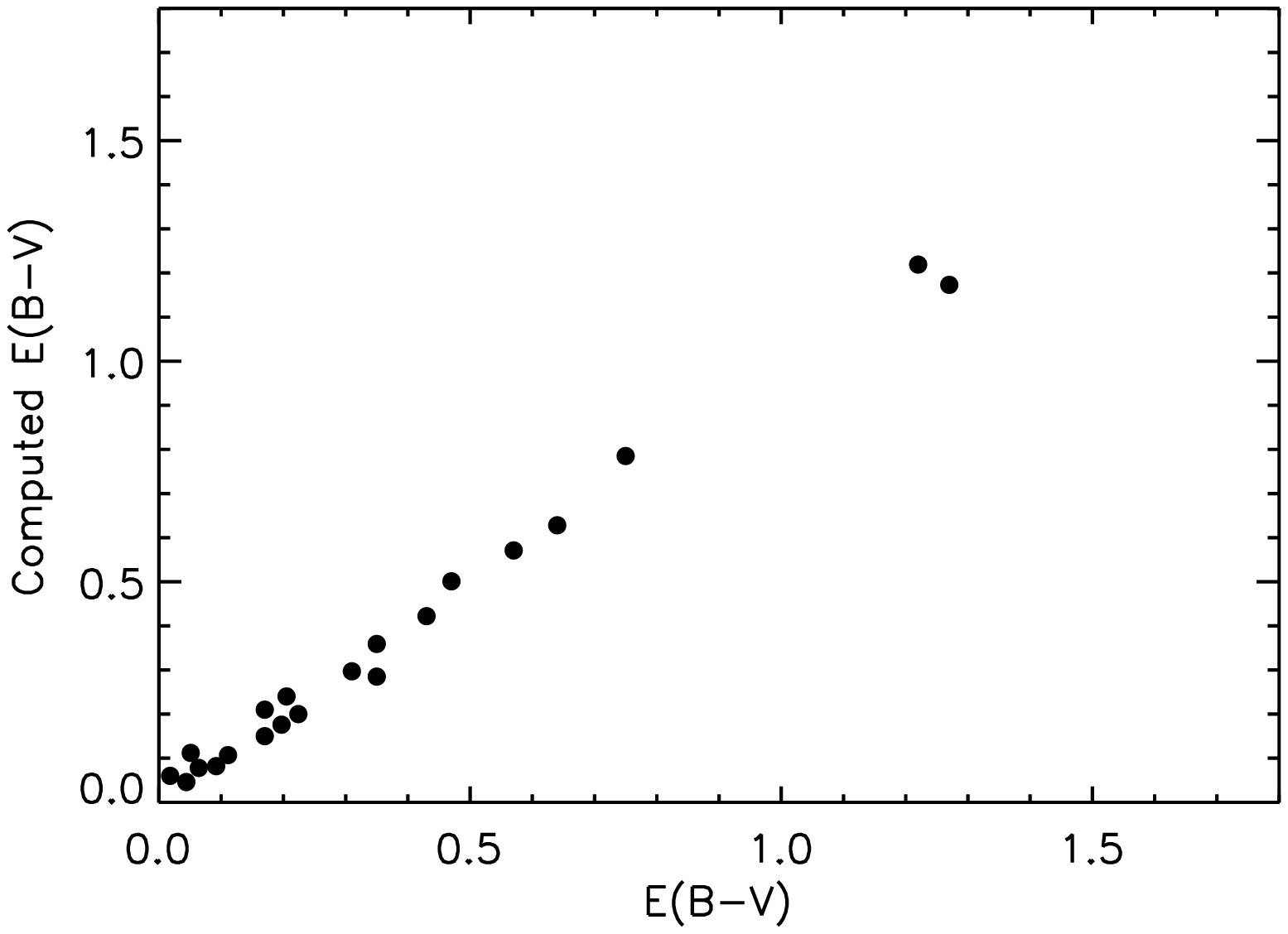}
\end{center}
\caption{A comparison of the output from the Cepheid relations developed in section \ref{framework} with literature values.}
\label{linearrel}
\end{figure}

A distance relation applicable to Cepheid variables is formulated here, motivated by the work of \citet{op83,op88}. Consider the canonical distance modulus equation:
\begin{equation}
\label{eqn1}
5\log{d}=V-A_V-M_V+5 \;,
\end{equation}
where $A_V=R_V\times E_{B-V}$ and $E_{B-V}=\left(B-V\right)-\left(B-V\right)_0$. The standard period-luminosity-colour (PLC) relation for Cepheids can be expressed as:
\begin{eqnarray}
\nonumber
M_V=a \log{P} + b (B-V)_0 + c \;.
\end{eqnarray}
Equation (\ref{eqn1}) can therefore be rewritten as:
\begin{eqnarray}
\nonumber
5\log{d}=V- a \log{P} - b (B-V) + (b-R_V) E_{B-V} - c + 5 \;,
\end{eqnarray}
or:
\begin{equation}
\label{eqn2}
5\log{d}=V+ \alpha \log{P} + \beta (B-V) + \delta E_{B-V} + \gamma  \;. \\
\end{equation}
A calibrating set (Table 1) consisting of established cluster Cepheids and Cepheids with parallaxes measured recently with the HST \citep{be07} was used to determine the co-efficients in equation (\ref{eqn2}) that minimize the $\chi^2$ statistic, yielding an optimum solution given by:  
\begin{eqnarray}
\nonumber
5\log{d}=V+3.77 \log{P} - 2.40 (B-V) - E_{B-V}  + 7.03 \;.
\end{eqnarray}
The resulting relationship reproduces the distances and colour excesses for the calibrating set with formal average uncertainties of $\pm 3$\% (Fig. \ref{linearrel}). The true scatter applying to use of the relationship for individual Cepheids may be larger, given that the calibrating set consists primarily of large-amplitude Cepheids lying near the centre of the instability strip.

The colour excess term can also be characterized in terms of observable parameters as:
\begin{equation}
\label{eqn3}
E_{B-V}=\eta \log{P} + \lambda (V-J) + \phi \;,
\end{equation}
where ({\it V--J}) colour appears to be a viable surrogate for determining colour excess, although {\it H}- or {\it K}-band photometry could be substituted for {\it J}. A test of the colour excess relation using ({\it V--H}) and ({\it V--K}) produced slightly larger $\chi^2$ statistics than when ({\it V--J}) was used as the colour index, so the latter was adopted in the present study.

Suitable infrared and optical photometry for the calibrating Cepheids (Table \ref{calset}) was obtained from \citet{ls92}, \citet{gr99} and sources identified by \citet{fo07} (see discussion in their section 2.1), with {\it J}-band measures being standardized on the 2MASS system. The co-efficients of equation (\ref{eqn3}) that minimize the $\chi^2$ statistic are:
\begin{equation}
\label{eqn4}
E_{B-V}=-(0.270) \log{P} + (0.415) (V-J) - 0.255 \;.
\end{equation}
The co-efficients of the same equation that minimize the $\chi^2$ statistic for ({\it V--H}) and ({\it V--K}) are:
\begin{eqnarray}
\nonumber
E_{B-V}=-(0.33) \log{P} + (0.37) (V-H) - 0.27 \;,
\end{eqnarray}
and
\begin{eqnarray}
\nonumber
E_{B-V}=-(0.30) \log{P} + (0.34) (V-K) - 0.27 \;.
\end{eqnarray}

Equation (\ref{eqn4}) reproduces the reddenings for the calibrating Cepheids with an average uncertainty of $\pm0.03$ mag. (Fig. \ref{linearrel}), although the true scatter applying to use of the relationship for individual Cepheids will be larger. The relation makes use of {\it V} and {\it J}-band photometry (2MASS) that are widely available, and should provide a first order estimate in the absence of reddenings determined by means of $BVI_c$ photometry \citep{lc07,ls94}, spectroscopic analyses \citep{ko08}, or space reddenings \citep{lc07,tu08}. Alternatively, a reddening-free distance relation analagous to the Wesenheit function \citep{ma82,vb68} can be constructed by setting $\delta=0$ [equation (\ref{eqn3})], which results in a negligible increase in the $\chi^2$ statistic relative to the optimum solution in equation (\ref{eqn3}). A reanalysis of the co-efficients then yields:
\begin{eqnarray}
\nonumber
5\log{d}=V+ (4.42) \log{P} - (3.43) (B-V) + 7.15 \;.\\
\nonumber
5\log{d}=V+ (3.43) \log{P} - (2.58) (V-I) + 7.50 \;.\\
\nonumber
5\log{d}=V+ (3.30) \log{P} - (1.48) (V-J) + 7.63 \;.
\end{eqnarray}

Published results raise questions about the pulsation modes of the s-Cepheids EV Sct and QZ Nor \citep{cc85,mb86,bo01}. With the relationship derived here, the memberships of EV Sct in the cluster NGC 6664 \citep{me87} and QZ Nor in the cluster NGC 6087 \citep{cc85,me87}, as established by radial velocity measures, imply that EV Sct and QZ Nor are overtone pulsators. Otherwise, equation (\ref{eqn3}) with the assumption of fundamental mode pulsation results in anomalous luminosities for the Cepheids, namely values that differ from those resulting from implied cluster membership by several times the mean uncertainties. Such conclusions are sensitive, however, to the distances adopted for both clusters in Table \ref{calset}. 

\begin{figure*}
\begin{center}
\includegraphics[width=12cm]{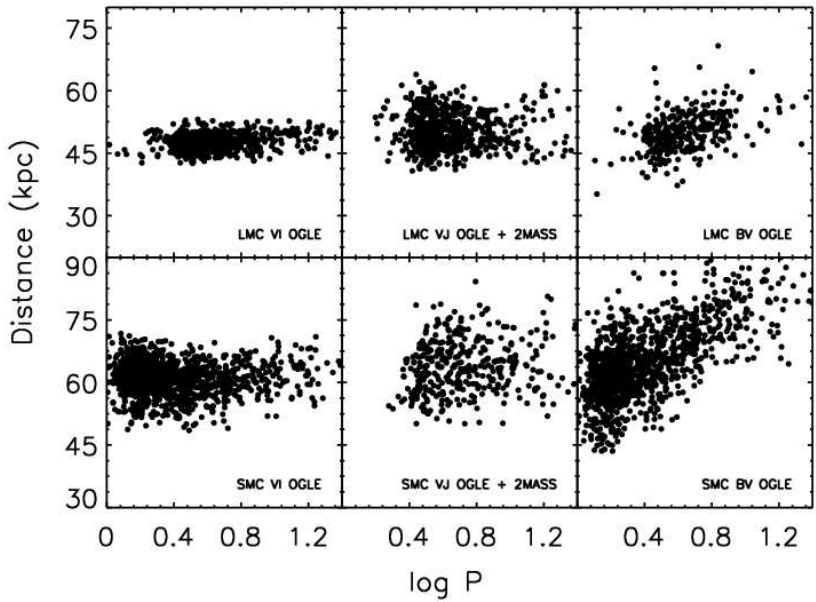}
\end{center}
\caption{Cepheid distance diagrams (CDD) constructed for the Magellanic Clouds. See text for details.}
\label{lmc}
\end{figure*}

\subsection{Determining $\langle J\rangle$}
Reliable estimates for a Cepheid's colour excess from equation (\ref{eqn4}) require the availability of precise mean {\it J}-band magnitudes. Infrared light curves are readily available for the calibrating set (Table \ref{calset}), but in most instances only single epoch 2MASS photometry exists. The derivation of mean magnitudes from single epoch observations is complicated by several issues. First, Cepheids undergo rapid period changes \citep{sz77,sz80,sz81,be94,be97,gl06,tu06}, so a significant time lapse between single epoch observations and those of the reference optical light curve can result in correspondingly large phase offsets. On the other hand, most Cepheids with periods between 5 and 10 days exhibit relatively slow period changes, as confirmed observationally in section \ref{cepheids}. Second, the mean magnitude deduced from single epoch observations is less certain because the morphological structure of the light curve can change between the optical and infrared (e.g., SV Vul); the former traces the temperature, and the latter the radius. Deriving colour excesses for stars lacking multiple observations is undoubtedly less precise, especially for large amplitude Cepheids and those exhibiting a significant, yet usually unknown, period change.

The mean {\it J}-magnitude $\langle J \rangle$ can be approximated by:
\begin{eqnarray} 
\nonumber
\langle J \rangle \simeq J_{\rm JD} - \left( \frac{\mid V(\phi_J)-V_{\rm max} \mid}{V_{\rm a}} - 0.5 \right) \times J_{\rm a} \;,
\end{eqnarray}
where $J_{\rm JD}$ is the magnitude for the single epoch observation, $V(\phi_J)$ is the visual magnitude at the same phase as $J_{\rm JD}$, $V_{\rm max}$ is the {\it V}-band magnitude of the star at maximum brightness, and $V_{\rm a}$ and $J_{\rm a}$ are the light amplitudes of the Cepheid in the visual and {\it J}-band, respectively. For the {\it V} to {\it J} amplitude relation outlined by \citet{we84} and \citet{so05} of $J_{\rm a}\simeq0.37\times V_{\rm a}$, the equation becomes:
\begin{equation} 
\label{eqn5}
\langle J \rangle \simeq J_{\rm JD} - \left( \frac{\mid V(\phi_J)-V_{\rm max} \mid}{V_{\rm a}} - 0.5 \right) \times 0.37 V_{\rm a} \;.
\end{equation}

A derivation of the mean magnitude for the cluster Cepheids DL Cas, CV Mon, QZ Nor, V340 Nor, and EV Sct, which are not saturated in the 2MASS survey and have been observed at a fairly recent epoch by ASAS, thereby minimizing the effects of period changes, yields an average difference of $\pm0.03$ mag. relative to mean {\it J}-band mangitudes found in the literature. That may be an optimistic estimate given that the light curves for Cepheids in the above sample are primarily sinusoidal and of small amplitude. In spite of the cited uncertainties for single epoch 2MASS observations ($\Delta J\simeq \pm0.03$ mag.) and the above considerations, equation (\ref{eqn5}) proves to be a satisfactory approximation for determining $\langle J \rangle$.

\subsection{Extragalactic Comparisons}
\label{extragal}
The distance to the LMC and SMC can be established by adopting the reddening-free relations highlighted earlier and utilizing {\it V} and {\it I} photometry from OGLE \citep{ud99}, {\it V} and {\it J} photometry from a combined set of OGLE and 2MASS data compiled by the authors (see Fig. \ref{lmc}),\footnote{The OGLE + 2MASS dataset for the LMC and SMC are available online at the Vizier database.} and {\it B} and {\it V} photometry from OGLE. The first two methods yield distance moduli to the LMC of $18.39\pm0.09$ and $18.49\pm0.19$. The distance moduli to the SMC derived from {\it VI} and {\it VJ} photometry are $18.93\pm0.14$ and $19.02\pm0.22$.

The diagrams constructed from reddening-free {\it VI} and {\it VJ} relations remain generally unbiased towards redder colours, but there is an obvious bias for the {\it BV} distances that may be attributed to line blanketing effects arising from metallicity differences among Milky Way, LMC and SMC Cepheids.  The effect is more pronounced for the SMC, an expected trend given that SMC Cepheids exhibit a lower metallicity than those of the LMC \citep[e.g.,][]{mo06}. The results confirm that the slope of the PL relation is not universal when based on {\it BV} photometry, while by comparison, the slopes for the {\it VJ} and {\it VI} relations seem relatively unaffected by metallicity. The derived distances are consistent with values found in the literature \citep[see][]{ls94,be02}), although distances cited in other studies are preferred since a putative zero-point metallicity correction was not addressed here. The {\it VJ} result establishes the viabilitiy of 2MASS photometry in such analyses in spite of the survey's single epoch observations. Indeed, the uncertainty in the {\it VJ} result could be reduced by approximating the mean $\langle J \rangle$-band magnitude according to the prescription described earlier or that described by \citet{so05}. Lastly, it is noted that biases towards redder colours may also result from standardization problems \citep[see Figure 14 of][]{za02}. 

The giant elliptical galaxy NGC 5128 hosts large numbers of Cepheids, and \citet{fe07} derived a distance of $3.1\pm0.1$ Mpc to the galaxy using the Wesenheit PL calibration (see Table 6 and section 5.3 in their paper). The distance to NGC 5128 established with the reddening-free {\it VI} relation formulated in section \ref{framework} is $3.1\pm0.5$ Mpc. Star C43 in Table 5 of \citet{fe07} is presumably not a Type I Cepheid. Similarly, \citet{ph98} derived a distance of $12.03\pm0.9$ Mpc to NGC 2090 from Cepheids, while the reddening-free {\it VI} relation given here yields a comparable distance of $11.8\pm1.3$ Mpc. A broader analysis including a larger sample of galaxies is needed to draw any meaningful conclusions, but to first order the results are in agreement.

\section{Potential Cluster Cepheids}
\label{cepheids}
\subsection{GSC 03729-01127}
\label{gsc}
\begin{figure}
\begin{center}
\includegraphics[width=8cm]{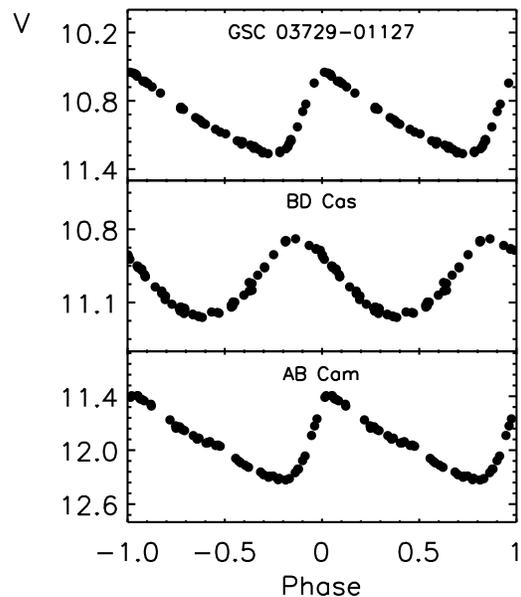}
\end{center}
\caption{The light-curves for the Cepheids GSC 03729-01127, BD Cas, and AB Cam, constructed from differential photometry obtained at the Abbey Ridge Observatory. Standard deviations across check stars in the Cepheid fields typically range from $\pm0.006$ to $\pm0.008$ mag.}
\label{photometry}
\end{figure}

The Cepheid-like variations of GSC 03729-01127 were first noted by Mike Sallman while inspecting TASS observations, a discovery that led to its identification with an earlier entry in the NSVS \citep[object 1973907,][]{wo04}. Preliminary analysis of the TASS data produced a period of $P\simeq5.074$ days, a value closely approximated by our observations, which yield $P=5.065\pm0.008$ days. A period analysis of the photometry was carried out in the \textsc{peranso} software environment \citep{va07} using the algorithms \textsc{anova} \citep{sc96}, \textsc{falc} \citep{ha89}, and \textsc{cleanest} \citep{fo95}. The phased light curve (Fig. \ref{photometry}) has an amplitude of $\Delta V\simeq0.72$ mag. ($\Delta B\simeq1.09$ mag. in the blue), and displays a typical Cepheid signature with a rapid rise from minimum to maximum. 

Spectroscopy confirms that the variable is indeed a Cepheid, displaying spectral variations from F6 Ib to G1 Ib over its cycle. All-sky {\it BV} photometry for GSC 03729-01127 was obtained on several nights, with extinction co-efficients derived using techniques outlined by \citet{he98} and \citet{wa06}. The photometry was standardized to the Johnson system using stars in the nearby open cluster NGC 225 \citep{ho61}. The mean magnitude and colour are $\langle V\rangle \simeq 10.90$ and $\langle B-V\rangle \simeq 1.67$, which, with the formulation of section \ref{framework}, result in an estimated distance of $d=1230\pm120$ pc and a colour excess of $E_{B-V}=1.05\pm0.05$. The Cepheid is assumed to be pulsating in the fundamental mode, as inferred from the morphological structure of its light curve \citep[see][]{be95,bs98,we95}.

\begin{figure*}
\begin{center}
\includegraphics[width=12cm]{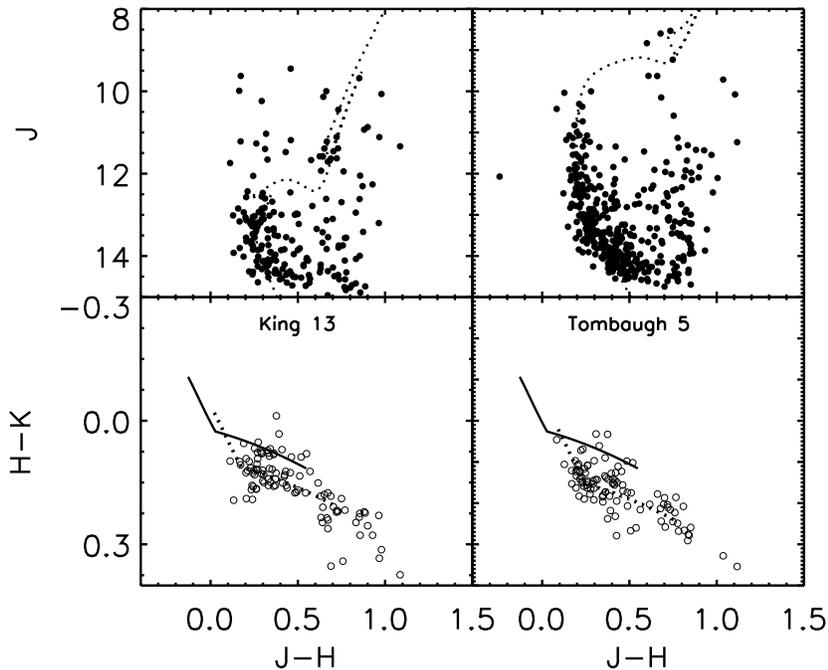}
\end{center}
\caption{Colour-colour and colour-magnitude diagrams for King 13 (left) and Tombaugh 5 (right) constructed from 2MASS data. Clearly visible sequences of AF-type stars (King 13) and late B-type stars (Tombaugh 5) produce reddenings of $E_{J-H}=0.14\pm0.02$ ($E_{B-V}\simeq0.52$) and $E_{J-H}=0.22\pm0.02$ ($E_{B-V}\simeq0.81$), respectively. King 13 and Tombaugh 5 have inferred ages of $\log \tau =9.0\pm0.2$ and $\log \tau =8.35\pm0.15$, and distances of $d=2.55\pm0.50$ kpc (see text) and $d=1.66\pm0.20$ kpc, respectively. A reddening relationship of $E_{J-H}=1.72\times E_{H-K}$ was adopted from \citet{bb05} and \citet{du02}.}
\label{ki13to5ccmd}
\end{figure*}

The Cepheid lies $22\arcmin$ from the core of Tombaugh 5 \citep{to41}, an open cluster estimated to be $\sim1.8$ kpc distant \citep{re54}. \citet{mn07} provide a more recent estimate of $d=1.33\pm0.33$ kpc, which is smaller than $d=1.75\pm0.15$ kpc derived by \citet{la04} from multi-band photometry. The data of both groups imply a reddening of $E_{B-V}\simeq0.8$. An analysis of 2MASS photometry for the field, fitted with a solar isochrone (Fig. \ref{ki13to5ccmd}) from the Padova Database of Stellar isochrones \citep{bo04}, implies a cluster age of $\log \tau =8.35\pm0.15$, a distance of $d=1.66\pm0.20$ kpc, and a reddening of $E_{J-H}=0.22\pm0.02$ ($E_{B-V}\simeq0.81$ according to the relations established in section \ref{oirelation}).
 
Star counts were made for Tombaugh 5 from 2MASS data, relative to a cluster centre at 03:47:56, +59:04:59 (J2000) found from strip counts on the Palomar survey E-plate of the field. The data (Fig. \ref{starcounts}) imply a nuclear radius for the cluster of $r_{\rm n} \simeq 8$ arcminutes, and a coronal radius of $R_{\rm c} \simeq 23$ arcminutes, in the notation of \citet{kh69}. GSC 03729-01127 lies close to the cluster's tidal limit, which raises questions about its possible association with Tombaugh 5. The progenitor mass of GSC 03729-01127 can be estimated from its pulsation period as $\sim 4 M_{\sun}$ \citep[see][]{tu96}, consistent with the implied turnoff mass for Tombaugh 5 and a possible physical association.

\setcounter{table}{1}
\begin{table}
\caption{Likely Evolved B-type Members of Tombaugh 5.}
\label{to5members}
\begin{center}
\begin{tabular}{cc}
\hline
2MASS Designation &$J$ \\ 
\hline
03473515$+$5907588 &10.307 \\
03482167$+$5903410 &11.607 \\
03481681$+$5901295 &11.321 \\
03471767$+$5904333 &11.542 \\
03473265$+$5901398 &11.589 \\
03481650$+$5905219 &12.018 \\
\hline
\end{tabular}
\end{center}
\end{table}

Although the star count evidence is only marginally consistent with membership of GSC 03729-01127 in Tombaugh 5, the similarity of the distance estimates and evolutionary ages for Cepheid and cluster are sufficient to warrant follow-up radial velocity measures, which are needed before reaching any conclusions regarding membership. A list of likely cluster members has been tabulated for such a purpose (Table \ref{to5members}), with membership inferred from a correlation between a star's position in the 2MASS colour-colour and colour-magnitude diagrams. 

The Cepheid's evolutionary history was examined through O--C analysis (Fig. \ref{ocdiagrams}, Table \ref{gscoctable}), based primarily upon photographic light curves obtained from examination of archival images in the Harvard College Observatory Photographic Plate Collection, supplemented by CCD observations from NSVS, TASS, ASAS, the ARO, and data from a continuing program of Cepheid monitoring at the Burke-Gaffney Observatory of Saint Mary's University \citep{te98,tu99,tu05}. An adopted ephemeris of:
\begin{eqnarray}
\nonumber
{\rm JD}_{\rm max}=2453297.1935 + 5.06535 E \;,
\end{eqnarray}
where {\it E} is the number of elapsed cycles, was found to fit the observations reasonably well over the last century, and was used to phase the data. GSC 03729-01127 appears to be undergoing a gradual period increase of $+0.272 \pm0.088$ s yr$^{-1}$, confirming theoretical predictions on rates of period change for $P\simeq 5^{\rm d}$ Cepheids as outlined by \citet{tu06}. The rate of period change indicates that the Cepheid is evolving towards the red edge of the instability strip in the third or fifth crossing \citep{tu06}. There is some ambiguity, given the Cepheid's small pulsation amplitude, but that is resolved from the reddening derived here. GSC 03729-01127 appears to lie on the cool (red) edge of the instability strip, in which case its rate of period increase and small pulsation amplitude imply that it is in the fifth crossing of the strip.

\setcounter{table}{2}
\begin{table}
\caption{O--C Data for GSC 03729-01127.}
\label{gscoctable}
\begin{tabular}{cccccc}
\hline
JD$_{\rm max}$ &Cycles &O--C &Data &Weight &Source \\
& &(days) &Points \\ 
\hline
2411726.028 &$-8207$ &$+0.162$ &13 &1.0 &(1) \\	
2412480.634 &$-8058$ &$+0.031$ &13 &1.0 &(1) \\
2413959.564 &$-7766$ &$-0.121$ &25 &1.0 &(1) \\
2416127.206 &$-7338$ &$-0.450$ &11 &1.0 &(1) \\
2416816.473 &$-7202$ &$-0.070$ &22 &1.0 &(1) \\
2418492.862 &$-6871$ &$-0.312$ &19 &1.0 &(1) \\
2420417.567 &$-6491$ &$-0.440$ &24 &1.0 &(1) \\
2422894.642 &$-6002$ &$-0.321$ &8 &1.0 &(1) \\
2425720.892 &$-5444$ &$-0.536$ &8 &1.0 &(1) \\
2426855.472 &$-5220$ &$-0.595$ &28 &1.0 &(1) \\
2427225.308 &$-5147$ &$-0.529$ &11 &1.0 &(1) \\
2428841.526 &$-4828$ &$-0.157$ &27 &1.0 &(1) \\
2430563.605 &$-4488$ &$-0.298$ &15 &1.0 &(1) \\
2432463.383 &$-4113$ &$-0.026$ &14 &1.0 &(1) \\
2440537.114 &$-2519$ &$-0.463$ &13 &1.0 &(1) \\
2442710.182 &$-2090$ &$-0.430$ &27 &1.0 &(1) \\
2444584.756 &$-1720$ &$-0.036$ &24 &1.0 &(1) \\
2446534.863 &$-1335$ &$-0.089$ &34 &1.0 &(1) \\
2451478.579 &$-359$ &$-0.154$ &69 &3.0 &(2) \\
2453284.296 &$-3$ &$-0.049$ &94 &3.0 &(3) \\
2453808.799 &$+101$ &$+0.005$ &18 &3.0 &(4) \\
2454087.385 &$+156$ &$-0.003$ &49 &3.0 &(5) \\
\hline
\end{tabular}
Data sources: (1) Harvard Collection, (2) NSVS, \citet{wo04}, (3) TASS, \citet{dr06}, (4) Burke-Gaffney Observatory, (5) Abbey Ridge Observatory.
\end{table}

\setcounter{table}{3}
\begin{table}
\caption{New O--C Data for BD Cas.}
\label{bdcastable}
\begin{tabular}{cccccc}
\hline
JD$_{\rm max}$ &Cycles &O--C &Data &Weight &Source \\
& &(days) &Points \\ 
\hline
2412584.617 &$-8038$ &$-1.964$ &9 &1.0 &(1) \\
2412952.110 &$-7937$ &$-1.497$ &7 &1.0 &(1) \\
2413325.649 &$-7835$ &$-1.647$ &7 &1.0 &(1) \\
2415070.128 &$-7357$ &$-1.687$ &8 &1.0 &(1) \\
2415558.129 &$-7223$ &$-1.818$ &12 &1.0 &(1) \\
2415922.588 &$-7124$ &$-1.844$ &11 &1.0 &(1) \\
2416287.608 &$-7024$ &$-1.493$ &11 &1.0 &(1) \\
2416670.007 &$-6919$ &$-1.851$ &20 &1.0 &(1) \\
2417044.552 &$-6816$ &$-1.705$ &20 &1.0 &(1) \\
2417778.542 &$-6615$ &$-1.232$ &40 &1.0 &(1) \\
2419650.202 &$-6103$ &$-0.912$ &35 &1.0 &(1) \\
2421893.150 &$-5488$ &$-1.128$ &30 &1.0 &(1) \\
2423454.303 &$-5061$ &$-0.808$ &17 &1.0 &(1) \\
2424225.733 &$-4849$ &$-0.442$ &7 &1.0 &(1) \\
2425372.119 &$-4535$ &$-0.825$ &22 &1.0 &(1) \\
2425744.076 &$-4433$ &$-0.350$ &8 &1.0 &(1) \\
2426109.087 &$-4333$ &$-0.654$ &10 &1.0 &(1) \\
2427007.120 &$-4088$ &$-0.675$ &13 &1.0 &(1) \\
2429772.037 &$-3330$ &$+0.083$ &49 &1.0 &(1) \\
2431560.580 &$-2840$ &$+0.003$ &43 &1.0 &(1) \\
2433312.640 &$-2360$ &$-0.183$ &21 &1.0 &(1) \\
2440495.195 &$-393$ &$-0.144$ &19 &1.0 &(1) \\
2441932.979 &$+0$ &$-0.277$ &21 &2.0 &(2) \\
2442829.249 &$+245$ &$-0.294$ &66 &1.0 &(1) \\
2445089.520 &$+865$ &$-0.541$ &... &1.5 &(3) \\
2446379.817 &$+1218$ &$-0.509$ &100 &1.0 &(1) \\
2447482.216 &$+1520$ &$-0.431$ &14 &2.0 &(4) \\
2448218.562 &$+1721$ &$-0.438$ &98 &2.0 &(5) \\
2448797.229 &$+1880$ &$-0.442$ &28 &2.0 &(5) \\
2453086.229 &$+3055$ &$-0.500$ &55 &3.0 &(6) \\
2453858.027 &$+3266$ &$-0.506$ &17 &3.0 &(7) \\
2454047.561 &$+3318$ &$-0.529$ &56 &3.0 &(8) \\
\hline
\end{tabular}
Data sources: (1) Harvard Collection, (2) \citet{sz77,sz83}, (3) \citet{bu86}, (4) \citet{sc91}, (5) Hipparcos, \citet{pe97}, (6) TASS, \citet{dr06}, (7) Burke-Gaffney Observatory, (8) Abbey Ridge Observatory.
\end{table} 

\subsection{BD Cas}
BD Cas is a small-amplitude, sinusoidal, $3^{\rm d}.65$ Cepheid ($\Delta V\simeq 0.33$, Fig. \ref{photometry}) discovered by \citet{be31}. Increased sampling and precision photometry indicate that the variations are not purely sinusoidal (Fig. \ref{photometry}), although the star could still qualify as an s-Cepheid. There is some uncertainty regarding the variable's pulsation mode and population type \citep[][see references and discussion therein]{ki99}. \citet{ki99} examined the structure of the Cepheid's radial velocity curve in Fourier space using \textsc{coravel} measures and data from \citet{go96} and noted that the values of $A_1$ and $R_{21}$ are consistent with overtone pulsation, whereas $\phi_{21}$ is not. They argued that the analysis was hampered by large uncertainties because of undersampling, with further observations needed to constrain the pulsation mode unambiguously.

A Fourier analysis of the new ARO photometry yields an amplitude ratio of $R_{21}=0.113\pm0.010$ and $\phi_{21}=3.50\pm0.09$, which imply overtone pulsation by the criteria of \citet{bs98} and \citet{za05}. The Cepheid's near-solar metallicity of [Fe/H]$=-0.07$ \citep{an02} and Galactic location ($\ell,\textit{b}=118,-1$) are consistent with other Population I variables, which suggests that BD Cas is a classical Cepheid pulsating in the first overtone. The Cepheid's spectral type exhibits only small variations, from F6 II to F7 Ib. It should also be noted that a viable framework for discriminating a Cepheid's population type by means of Fourier analysis has yet to emerge \citep{fe99}.

BD Cas lies $20\arcmin$ from the open cluster King 13 \citep{ki49} and $\sim16\arcmin$ from the suspected cluster Czernik 1 \citep{cz66}. \citet{ml79} examined King 13 using photographic {\it UBV} photometry, and derived a distance of $d=1730\pm200$ pc and a reddening of $E_{B-V}=0.38$. \citet{sb07} obtained a larger reddening of $E_{B-V}=0.82\pm0.02$, a larger distance of $d=3100\pm300$ pc, and a cluster age of $\log \tau=8.5$ from CCD photometry, while \citet{mn07} obtained an age of $\log \tau=8.4$, a distance of $d=3670\pm1300$ pc, and a reddening of $E_{B-V}=0.86\pm0.12$ by similar means.  
An analysis of 2MASS photometry for the field, fitted with an isochrone from the Padova Database of Stellar isochrones \citep{bo04}, results in a distance of $d=2550\pm500$ pc and a reddening of $E_{J-H}=0.15\pm0.02$ ($E_{B-V}\simeq0.56$ according to the relations derived in section \ref{oirelation}). The colour-colour and colour-magnitude diagrams are dominated by AF dwarfs (Fig. \ref{ki13to5ccmd}) and the reddening solution is well established. Moreover, the structure of the main-sequence turnoff and the clustering of red giant members (Fig. \ref{ki13to5ccmd}) indicates an age near $\log \tau =9.0\pm0.15$. The distance estimate should be viewed cautiously because it is tied to stars lying near the limiting magnitude of the 2MASS survey. The age and reddening estimates are tied more directly to the colours of main sequence stars, and are consequently more reliable.

\begin{figure}
\begin{center}
\includegraphics[width=7cm]{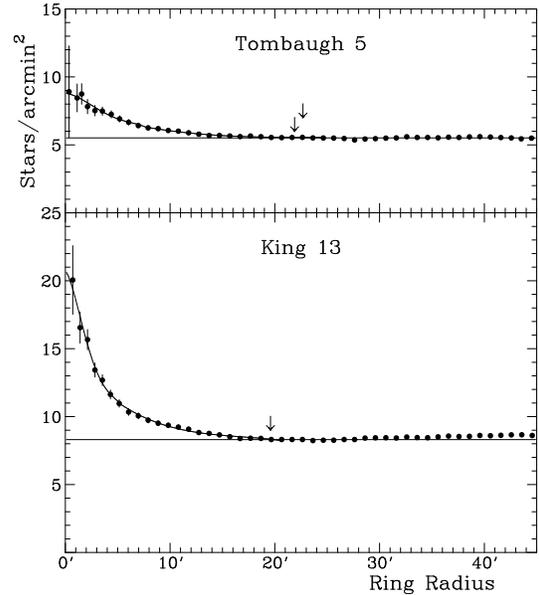}
\end{center}
\caption{Star counts for the open clusters Tombaugh 5 and King 13, compiled from 2MASS observations. Arrows indicate the locations of GSC 03729-01127 and AB Cam (inner and outer points, upper figure) and BD Cas (lower figure) relative to their respective cluster centres.}
\label{starcounts}
\end{figure}

Star counts were made for King 13 from 2MASS data, relative to a cluster centre at 00:10:16.47, +61:11:29.2 (J2000) found from strip counts on the Palomar survey E-plate of the field. The data (Fig. \ref{starcounts}) imply a nuclear radius for the cluster of $r_{\rm n} \simeq 6$ arcminutes, and a coronal radius of $R_{\rm c} \simeq 20$ arcminutes. BD Cas therefore lies very close to the cluster's tidal limit. With the Cepheid relationship formulated in section \ref{framework} and pertinent photometric parameters for BD Cas \citep{sz77}, the Cepheid has an estimated distance of $d=1520\pm150$ pc and a reddening of $E_{B-V}=0.99\pm0.05$, which places it well foreground to the cluster, but curiously with a larger reddening. The large reddening for BD Cas is confirmed by an independent spectroscopic reddening estimate of $E_{B-V}=1.01$ \citep{ko08}. A possible physical association between the cluster and Cepheid is further negated by the cluster's age, which implies a main-sequence turnoff mass near 2 $M_{\sun}$. There are no other known Population I cluster Cepheids associated with a such a correspondingly old cluster \citep{tu96}. 

\citet{ly95} has questioned the existence of the open cluster Czernik 1, which was suggested as an alternate parent cluster for BD Cas by \citet{ts66}. Our analysis of 2MASS photometry and limited {\it BV} photometry for the cluster is inconclusive in that regard, so a possible connection with Czernik 1 remains unproven.

The star's evolutionary history was examined through O--C analysis (Fig. \ref{ocdiagrams}) using light curves constructed from data obtained from archival plates in the Harvard collection, data compiled by \citet{sz77,sz83}, and more recent photoelectric and CCD photometry (Table \ref{bdcastable}). An ephemeris given by:
\begin{eqnarray}
\nonumber
{\rm JD}_{\rm max}=2441932.0320 + 3.65090 E \;,
\end{eqnarray}
was found suitable for phasing the data. BD Cas is undergoing a period decrease amounting to $-0.698 \pm0.048$ s yr$^{-1}$ (Fig. \ref{ocdiagrams}), which, in conjunction with its derived reddening and inferred overtone status, implies a second crossing of the instability strip \citep{tu06} and a star lying towards the hot (blue) edge. The rate of period change was derived from a polynomial fit (see Fig. \ref{ocdiagrams}), but there are deviations in the O--C data derived from recent photoelectric and CCD photometry that may indicate non-evolutionary effects. The trends are unlikely to arise from binarity, since that would imply an unrealistically large mass for the companion. Yet binarity is not precluded, and is common among Cepheids \citep{sz95,sz03}. The trends observed here and in the O--C diagrams of other Cepheids have yet to be explained satisfactorily, and remain an active and interesting area of research. Whether the superposed variations are small, as in the Cepheid RT Aur \citep{tu07}, or arise from random fluctuations in pulsation period \citep[e.g., SV Vul,][]{tu04}, the mechanism is yet to be established.  

\subsection{AB Cam}
AB Cam is a large-amplitude, $5^{\rm d}.79$ Cepheid ($\Delta V\simeq0.94$, Fig. \ref{photometry}) discovered by \citet{mo34}. The distance to AB Cam can be estimated from Johnson system photometry \citep{be00} with the Cepheid relationship formulated in section \ref{framework}. The implied distance is $d=4470\pm450$ pc with a reddening of $E_{B-V}=0.59\pm0.05$. The object's light-curve morphology (Fig. \ref{photometry}) is consistent with fundamental mode pulsation \citep{bs98,be95,we95}, which was assumed for estimating the distance. \citet{vb57} and \citet{ts66} examined the field of AB Cam in search of a host cluster with null results. Despite the Cepheid's apparent angular proximity to Tombaugh 5, the two are not physically related (see Fig. \ref{starcounts}).

\begin{figure}
\begin{center}
\includegraphics[width=8cm]{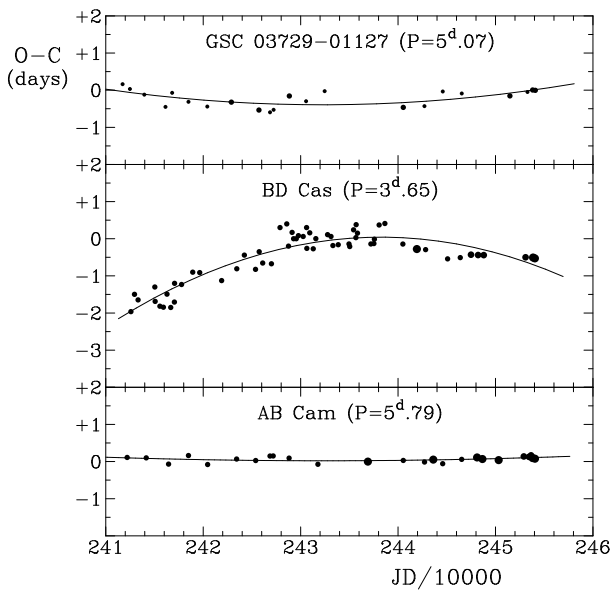}
\end{center}
\caption{The O-C diagrams for GSC 03729-01127, BD Cas, and AB Cam (from top to bottom), including observations compiled by \citet{sz77,sz83}. The size of each data point is proportional to the weight assigned in the analysis, and the parabola in each case represents a regression fit.}
\label{ocdiagrams}
\end{figure}

The star's evolutionary history was examined through O--C analysis (Fig. \ref{ocdiagrams}) using photographic light curves constructed with magnitude estimates from archival plates in the Harvard collection, supplemented by more recent photoelectric and CCD photometry (Table \ref{abcamtable}). An ephemeris given by:
\begin{eqnarray}
\nonumber
{\rm JD}_{\rm max}=2437400.6380 + 5.78745 E \;,
\end{eqnarray}
proved suitable for phasing the data. AB Cam appears to be undergoing a very gradual increase of $+0.069 \pm0.025$ s yr$^{-1}$. The rate is again consistent with observed rates of period change for $P\simeq 6^{\rm d}$ Cepheids \citep{tu06}, and indicates, in conjunction with its large pulsation amplitude, a Cepheid in the third crossing of the instability strip lying towards the cool (red) edge. The object may be of added importance given that it falls near the lower bound for an undersampled locus of short-period third crossers in the rate of period change diagram \citep{tu06}.

\setcounter{table}{4}
\begin{table}
\caption{O--C Data for AB Cam.}
\label{abcamtable}
\begin{tabular}{cccccc}
\hline
JD$_{\rm max}$ &Cycles &O--C &Data &Weight &Source \\
& &(days) &Points \\ 
\hline
2412196.403 &$-4355$ &$+0.110$ &16 &1.0 &(1) \\
2414158.338 &$-4016$ &$+0.099$ &22 &1.0 &(1) \\
2416444.214 &$-3621$ &$-0.068$ &28 &1.0 &(1) \\
2418487.415 &$-3268$ &$+0.163$ &15 &1.0 &(1) \\
2420454.905 &$-2928$ &$-0.079$ &24 &1.0 &(1) \\
2423429.801 &$-2414$ &$+0.068$ &13 &1.0 &(1) \\
2425391.706 &$-2075$ &$+0.027$ &16 &1.0 &(1) \\
2426856.054 &$-1822$ &$+0.149$ &26 &1.0 &(1) \\
2427185.941 &$-1765$ &$+0.152$ &20 &1.0 &(1) \\
2428817.943 &$-1483$ &$+0.093$ &26 &1.0 &(1) \\
2431769.375 &$-973$ &$-0.074$ &92 &1.0 &(1) \\
2436902.916 &$-86$ &$-0.002$ &28 &3.0 &(2) \\
2440531.679 &$+541$ &$+0.030$ &13 &1.0 &(1) \\
2442713.505 &$+918$ &$-0.012$ &26 &1.0 &(1) \\
2443610.625 &$+1073$ &$+0.053$ &14 &3.0 &(3) \\
2444577.019 &$+1240$ &$-0.057$ &24 &1.0 &(1) \\
2446527.506 &$+1577$ &$+0.059$ &34 &1.0 &(1) \\
2448107.531 &$+1850$ &$+0.111$ &13 &3.0 &(4) \\
2448657.297 &$+1945$ &$+0.069$ &9 &3.0 &(5) \\
2450324.052 &$+2233$ &$+0.038$ &13 &3.0 &(6) \\
2452905.354 &$+2679$ &$+0.137$ &18 &2.0 &(7) \\
2453432.000 &$+2770$ &$+0.126$ &26 &2.0 &(7) \\
2453651.965 &$+2808$ &$+0.167$ &19 &2.0 &(7) \\
2453802.368 &$+2834$ &$+0.096$ &12 &3.0 &(8) \\
2454051.208 &$+2877$ &$+0.076$ &45 &3.0 &(9) \\
\hline
\end{tabular}
Data sources: (1) Harvard Collection, (2) \citet{ba62}, (3) \citet{ha80}, (4) \citet{be92}, (5) \citet{ss96}, (6) \citet{be98}, (7) TASS, \citet{dr06}, (8) Burke-Gaffney Observatory, (9) Abbey Ridge Observatory.
\end{table} 

\section{Optical to Infrared Relations}
\label{oirelation}
Distance estimates for the open clusters studied in the preceeding section were established using the following expression:
\begin{displaymath}
5 \log{d}= [J-M_J] - E_{J-H}\times R_j +5\;.
\end{displaymath}

The infrared colour excess $E_{J-H}$ and the distance modulus {\it J--M}$_J$ can be derived by simultaneously fitting an intrinsic relation (Turner, unpublished) and isochrone \citep{bo04} to the stars in a colour-colour and colour-magnitude diagram (see Fig. \ref{ki13to5ccmd}). A ratio of total to selective extinction of $R_J=2.60$ was adopted \citep{bb05,du02}, and the following relationship between reddening in the infrared to that in the optical was used to permit direct comparison of cluster reddenings derived from 2MASS photometry with more frequently cited optical results found in the literature:
\begin{equation}
\label{eqn7}
E_{J-H}=(0.27\pm0.03) \times E_{B-V}\;.
\end{equation}

Equation (\ref{eqn7}) was established from the calibrating set of open clusters in Table \ref{calset}, and compares satisfactorily with relationships cited by \citet{ls93}, \citet{bb05} and \citet{du02}.  The co-efficient derived here is smaller, however, and may be indicative of our visual fitting biases rather than a global relationship. We refer the reader to \citet{ls93} for a more rigorous discussion of the correlation.

\section{Discussion}
The relationships highlighted in section 3 yield reliable parameters when investigating short period Cepheids ($P\le11^{\rm d}$). Parameters determined for longer period Cepheids from such relations are less certain, primarily because of an absence of mid-to-long period calibrators needed to identify a unique set of co-efficients consistent over a broad period baseline. At present $\ell$ Car is the only established long-period calibrator (parallax) used in deriving the co-efficients. A further drawback of the analysis rests in the adopted parameters for the calibrating clusters, which exhibit an unsatisfactory amount of scatter in the literature and more recent analyses \citep[e.g.,][]{an07,ho03}. Refining distance estimates to the calibrating set of clusters by means of deep CCD photometry, analagous to the impressive results from the CFHT Open Cluster Survey \citep{ka01a,ka01b}, is a priority in moving forward.  The framework is also tied to the HST sample of Cepheids with parallaxes and field reddenings established by \citet{be07}.  It is noted that the parallax measures for RT Aur and Y Sge differ significantly between HST and Hipparcos \citep[][see their table 1]{vl07}.  

The framework outlined here should permit an efficient investigation of suspected cluster Cepheids \citep{tu02}, including objects uncovered by cross correlations between newly discovered Cepheids in the ASAS and TASS with open cluster databases (i.e., WebDA). A potential goal is an expansion of the sample of cluster Cepheids, with particular emphasis on long period cluster Cepheids. Of equal importance, however, is the task of purging line-of-sight coincidences from current lists promulgating the literature, something that is particularly acute given that high-precision data are needed to address the question of the universality of the PL relation.  

Four longer term objectives exist. First is to use the new relations to determine Galactic parameters and map interstellar extinction. Second, is to establish mean photometry for an entire calibrating set. Third, with regard to the universality of the PL relation and establishing long-period Cepheid calibrators, realistically it will be the highly anticipated results from the GAIA mission \citep{ct06}, a next generation follow-up to the Hipparcos mission, that will provide the large and unbiased sample of Cepheid parallaxes needed to advance our knowledge of the field. In conjunction with a cleaned sample of cluster Cepheids, it should lead to a proper refinement of the relations outlined in section \ref{framework}, and, consequently, the realization of the outlined objectives.  Fourth is the longer term prospect of conducting extragalactic surveys using JWST \citep{ga06} to determine the distances and extinction to, and within [equation (\ref{eqn4})], higher redshift galaxies. The aperture size and infrared sensitivity of the telescope will permit deeper sampling of extragalactic Cepheids, especially since the variables are substantially brighter in the infrared than the optical and the diminishment in flux from reddening is comparitively less.  
  
\subsection*{ACKNOWLEDGEMENTS}
We are indebted to the following individuals and groups who helped facilitate the research: Alison Doane and the staff of the Harvard College Observatory Photographic Plate Stacks, Charles Bonatto for useful discussions on taking advantage of data from the 2MASS survey, Pascal Fouqu\'{e}, Laszlo Szabados and Leonid Berdnikov, whose comprehensive work on evolutionary trends in Cepheid variables was invaluable in our analysis, Arne Henden and the staff at the AAVSO, Dmitry Monin, Les Saddelmeyer, and the rest of the staff of the Dominion Astrophysical Observatory, Doug Welch who maintains the McMaster Cepheid Photometry and Radial Velocity Archive, the staff at la Centre de Donn\'{e}es astronomiques de Strasbourg, and Carolyn Stern Grant and the staff at the Astrophysics Data System (ADS). Reviews on Cepheids by Michael Feast, Donald Fernie, and Nick Allen were useful in the preparation of this work.   Lastly, we extend a special thanks to Sandra Hewitt for her exceptional kindness in accommodating visiting astronomers to the Harvard Plate Stacks.

\end{document}